\documentclass[journal=apchd5,manuscript=article,layout=twocolumn]{achemso} 
\usepackage{amsmath}
\usepackage{graphicx}
\usepackage{esint}
\usepackage{siunitx}
\usepackage{color}
\usepackage{ulem} 

\title{A telecom band single-photon source using a grafted carbon nanotube coupled to a fiber Fabry-Perot cavity in the Purcell regime.}

\author{Antoine Borel}
\author{Théo Habrant-Claude}
\author{Federico Rapisarda}
\affiliation{Laboratoire de Physique de l'ENS, Université PSL, CNRS, Sorbonne Université, Université Paris Cité, Paris, France}

\author{Jakob Reichel}
\affiliation{Laboratoire Kastler-Brossel, ENS, Sorbonne Université, Paris, France}

\author{Steeve Doorn}
\affiliation{Center for Integrated Nanotechnologies, Materials Physics and Applications Division, Los Alamos National Laboratory, Los Alamos, NM, USA}

\author{Christophe Voisin}
\author{Yannick Chassagneux}
\email{yannick.chassagneux@phys.ens.fr}
\affiliation{Laboratoire de Physique de l'ENS, Université PSL, CNRS, Sorbonne Université, Université Paris Cité, Paris, France}


\begin{document} 



\begin{abstract}
We report on the coupling of a reconfigurable high Q fiber micro-cavity to an organic color center grafted to a carbon nanotube for telecom wavelength emission of single photons in the Purcell regime. Using three complementary approaches we assess various figures of merit of this tunable single photon source and of the cavity quantum electrodynamical effects : the brightening of the emitter is obtained by comparison of the count rates of the very same emitter in free-space and cavity coupled regimes. We demonstrate a fiber coupled single-photon output rate up to 20 MHz at 1275~nm. Using time-resolved and saturation measurements, we determine independently the radiative quantum yield and the Purcell factor of the system with values up to 30 for the smallest mode volumes. Finally, we take advantage of the tuning capability of the cavity to measure the spectral profile of the brightness of the source which gives access to the vacuum Rabi splitting $g$ with values up to $25 \; \mu$eV.
\end{abstract}




\section{Introduction}
A range of quantum technologies, including quantum telecommunications or all optical quantum computing units rely on the availability of high quality single-photon sources which make up the ultimate flying quantum-bits. Among condensed-matter nano-emitters, self-assembled III-V quantum dots \cite{senellart2017high,Tomm2021,ward2005demand}, point defects in host matrices \cite{Albrecht2013,Dibos2018,Redjem2020} or 2D materials are promising candidates \cite{he2015single,Fournier2021}. However, few of these systems emit photons in the telecom bands (around 1.3 $\mu$m and 1.55 $\mu$m) where the powerful architecture of silicon based photonics and the fiber networks infrastructures could be exploited \cite{Cao2019,Hollenbach2020,Wang2018J}. Single-wall carbon nanotubes (SWCNTs) are among the most advanced systems in this respect \cite{He2018}, with the recent discovery of $\text{sp}^3$ diazonium grafting of organic color centers that unlocked the realization of room temperature, highly antibunched, telecom band emission of single-photons \cite{He2017a,Luo2019}. Nevertheless, the only way to harness the full power of such emitters is to efficiently couple them to photonic structures capable of controlling their emission properties through cavity quantum electrodynamics (cQED) effects and to funnel efficiently the photons into the working channel. Here, we demonstrate the efficient coupling in the Purcell regime ($\text{F}_{\text{P}}\sim$~30) of individual grafted carbon nanotubes to an open fiber Fabry-Perot cavity. We reach a single mode fiber-coupled stream of anti-bunched photons of \SI{2e7}{s^{-1}} at a wavelength of \SI{1.3}{\micro m}. Exploiting the unique versatility of our experimental setup, we were able to compare the luminescence properties of the same individual emitter in free-space (FS) and upon its coupling with a spectrally and spatially matched cavity. We could also vary continuously the key parameters of the cavity (such as the resonant wavelength or the mode volume) to carry an in-depth investigation of cQED effects. Using three independent approaches, we quantitatively assess the efficiency of the emitter/cavity coupling. We first compared the brightness of the same nano-emitter before and after the reversible coupling to the cavity, showing not only spectral filtering of the emission but also a strong brightening (x20). Furthermore, using time-resolved measurements we could directly observe the Purcell acceleration of the radiative recombination by comparing the lifetime of the same nano-emitter with or without the cavity and by tuning the mode volume of the latter. Finally, we were able to investigate an original feature of these mixed 0D/1D nano-emitters, by exploiting the strong acoustic phonon sidebands \cite{Galland2008,Vialla2014a} to tune the working wavelength of the single-photon source \cite{Jeantet2017a}. By analysing the variation of the single photon source efficiency as a function of detuning, we propose an third independent estimate of the strength of the cavity/emitter coupling. Finally, we discuss the strength and limitations of each of these approaches to qualify the relevance of each figure of merit with respect to applications.


\section{Samples and setup}
The sample consists in CoMoCAT (7,5) SWCNTs wrapped in a polymer (PFO-BPy) and dispersed in a toluene solution containing polystyrene (5 \% mass) before being spin-coated on a high reflectivity Bragg mirror, resulting in isolated SWCNTs in a 160~nm thick layer of polystyrene (PS). Carbon nanotubes were then functionnalized with 3,5 dichlorobenze diazonium salt by the dip doping method \cite{He2017a}. The 3,5 dichlorobenzene ion is covalently attached to the SWCNT surface by hybridization of $\text{sp}^2$ C-C bound into $\text{sp}^3$. Such hybridization  introduces a very localized 100-200 meV deep potential well for the center of mass of excitons, resulting in efficient trapping of the photo-generated excitons in the nanotube. This trapping leads to a red-shifted photo-luminescence and to a robust photon antibunching up to room temperature \cite{Ma2015a,He2017a}.

\begin{figure}[ht]
\includegraphics[width=8.6cm]{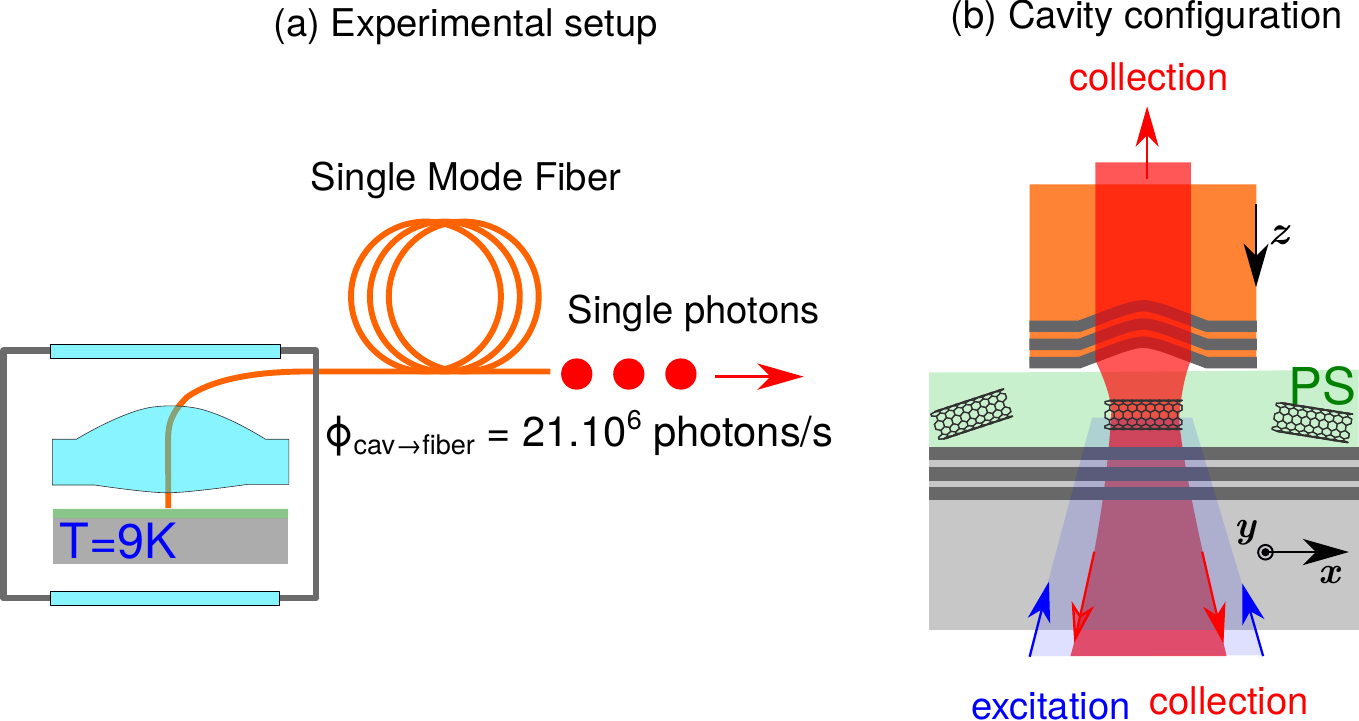} \caption{(a) Overview of the experimental setup producing a fibered stream of \SI{2.1e7}{} single-photons per second at a wavelength of 1.3 $\mu$m. (b) Zoom on the fibered micro-cavity with excitation and collection schemes.  \label{fig:setup} } 
\end{figure}

The experimental setup consists of a modified confocal microscope designed to observe the very same nano-emitter either in free space or coupled to a cavity \cite{Jeantet2016}. The cavity is depicted in Fig.~\ref{fig:setup}~(b). It consists of a flat Bragg mirror and a concave mirror engineered by CO$_2$ laser ablation at the apex of an optical fiber. The typical curvature radius is 10~$\mu$m while the cladding is pencil shaped by multiple laser impacts in order to access the shortest possible cavity length. The microscope objective is an aspheric lens at the centre of which a 300 µm diameter hole is drilled to pass the optical fiber. The flat mirror and the fiber apex were coated by similar dielectric stacks to reach a theoretical reflection coefficient $R\simeq 0.9995$ for the flat mirror and $R\simeq 0.9997$ for the fibered mirror in the wavelength range 1250 - 1500 nm.

The fiber-lens assembly system is mounted on an inertial nano-positionner for uniaxial displacement up to 2 mm on the z axis (Fig.~\ref{fig:setup}~(a,b)). This displacement is used to open and close the cavity at will (and hence to switch from the free-space to the cavity configurations), to adjust the mode volume (by changing the cavity length by $\lambda/2$ steps) and to finely tune the cavity  resonance to match each individual nano-emitter. In addition, a 1 µm course piezoelectric actuator is used for fast modulation of the cavity length. The Bragg flat mirror is clamped on two 2 mm range inertial nano-positionners for displacement in the (x,y) flat in order to adjust the spatial coupling of the nano-emitter to the cavity mode.  

Excitation of the nano-emitter was set to 1064 nm with either a Titanium:Sapphire cw laser (Spectra Physics 3900s) or with the seed of a Supercontinuum pulsed laser (NKT Photonics Fianium 4W, $\tau_{p}=8$ ps, $f_{rep}=$38.26 MHz). The photoluminescence (PL) is collected through the drilled aspheric lens in FS configuration, while in cavity configuration it is collected either through the  flat mirror or at the output port of the manufactured fiber. The PL is analyzed with a dispersive spectrometer and an InGaAs photodetector array for spectral measurements or with broadband fibered superconducting single photon detectors (Scontel SSPD, 32 ps jitter) for time-resolved measurements. 

\begin{figure}[ht]
\includegraphics[width=8.6cm]{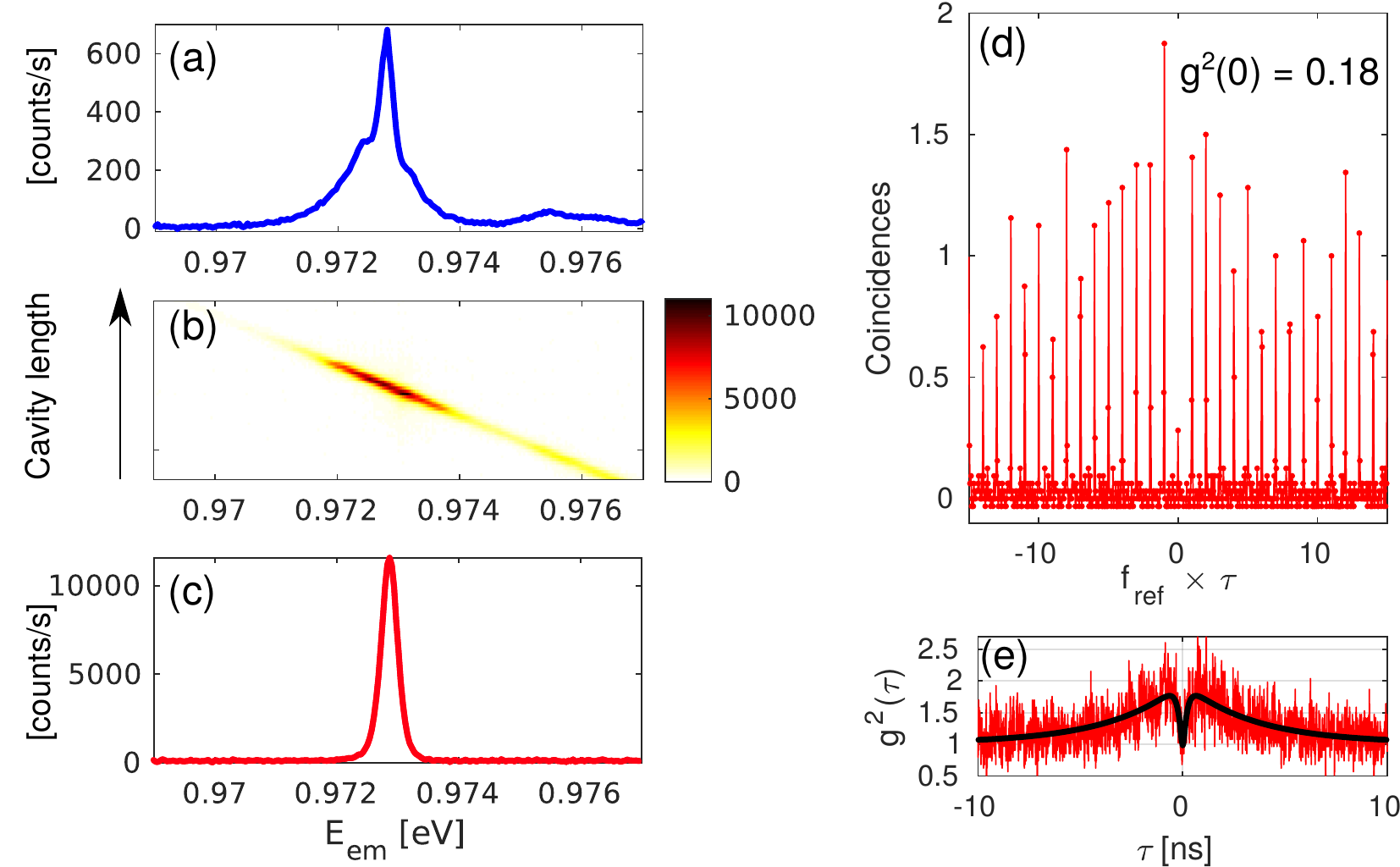} \caption{(a) Free space emission spectrum. (b) 2D plot of the emission spectrum recorded as a function of the cavity length. Color scale indicates the intensity. (c) Spectrum obtained at the spectral matching of the cavity mode with the emitter ZPL (Note that the cavity spectrum is broadened due to vibrations). Second order correlations in cavity (Fabry-Pérot mode p=6) under pulsed excitation, $g^2(0)=0.18$. (d), and under cw excitation, $g^2(0)=0.4$ (down to 0.36 once corrected by the response function), (e). A. clear bunching is observed under cw excitation and is attributed to intermittency in the emission with typical time scale of 10 ns. \label{fig:PL}}
\end{figure}

The typical low-temperature PL spectrum of an individual grafted nanotube is displayed in Fig.~\ref{fig:PL}~(a). In addition to the main line (hereafter called zero-phonon line (ZPL)), the PL spectra  consistently show well developed sidebands which result from the efficient coupling of the localized exciton to the 1D acoustic phonon bath of the CNT \cite{Galland2008,Vialla2014a}. The Debye-Waller factor (ratio of the ZPL to the integrated intensity) is typically 0.6-0.8. 
The intensities of the sidebands are directly related to the Bose-Einstein occupation probabilities of the phonon modes. Thus, at cryogenic temperature, the blue sideband (phonon absorption) is reduced with respect to the red one (phonon emission). In addition, the lines are broadened by pure dephasing and spectral diffusion. The latter is minimized by the use of a polystyrene matrix and by using a minimal excess energy in the photo-excitation process \cite{Walden-Newman2012}.  

Using a pair of fast SSPD and a correlation board, we performed a Hanbury-Brown and Twiss experiment to measure the intensity correlations of the photon stream, as shown in Fig.~\ref{fig:PL}~(d). The data show a clear anti-bunching behavior with $g^{(2)}(0) \le 0.2$ and a bunching behavior in the 10~ns range (better seen in the cw measurement shown in Fig.~\ref{fig:PL}~(e). This bunching possibly reflects the fast time-scale intermittency of the emission related to the existence of a dark triplet state, or a fast spectral diffusion exceeding the spectral width of the cavity.

\section{Brightening of the emitter through cQED effects}

In a high quality factor micro-cavity the spectral density of electromagnetic modes can be deeply modified, yielding an increase of the radiative decay rate (the so-called Purcell effect) when a nano-emitter is placed in the field anti-node \cite{purcell1946spontaneous} of a resonant mode. In a simple picture,  the coupling to the cavity opens a new decay channel for the emitter. At resonance, the total decay rate of the emitter reads \cite{auffeves2010} $\gamma_{\text{cav}}=\gamma_{\text{fs}} + DW F_\text{P}\gamma^\text{R} $, where $\gamma_{\text{fs}} = \gamma^{\text{NR}} + \gamma^\text{R}$ is the free-space decay rate with $\gamma^{\text{NR}}$ and $\gamma^\text{R}$ the non-radiative and radiative decay rates respectively. $F_\text{P}$ is the Purcell factor which is defined as the ratio of the radiative decay rate towards the resonant cavity mode to the FS radiative decay in the ZPL. The ratio of the emission in the ZPL to the entire spectrum is given by the Debye-Waller factor $DW$.  The Purcell factor reads \cite{auffeves2010} :
\begin{equation}
    F_\text{P}=\frac{3}{4\pi^2}\frac{(\lambda/n)^3}{V_{\text{eff}}}Q_{\text{eff}}\ , \label{eq:purcell_theo}
\end{equation}
with $V_{\text{eff}}=\frac{\int |nE|^2 d^3r}{|nE|^2_{\text{max}}}$  the effective mode volume and $Q_{\text{eff}}=(Q_{\text{cav}}^{-1}+Q_{\text{em}}^{-1})^{-1}$ the effective quality factor.

Grafted carbon nanotubes are quantum emitters that suffer from important non radiative decay channels. Even if room temperature quantum yields up to $\sim20$\% have been reported  \cite{Piao2013}, the low-temperature behavior is not documented. We show below that it remains very weak at the single emitter level so that non-radiative decay amounts for more than $90 \%$ of the total decay rate of our emitters in free space.
Thus, the Purcell effect, which modifies the radiative rate only, should  only weakly modify the total decay rate of the emitter on the one hand but on the other hand it should significantly increase its brightness. The ratio of the decay rates and emission flux in FS \textit{vs} cavity read : 
%
\begin{align}
    \frac{\phi_{\text{cav}}}{\phi_{\text{fs}}} &= \frac{DW F_{\text{P}}}{1+\eta_{QY}DW F_{\text{P}}}, & \text{(linear regime)} \\
    \frac{\phi_{\text{cav}}}{\phi_{\text{fs}}} &=DW F_{\text{P}}, \quad & \text{(saturation regime)} \label{eq:sat-rates}\\
   \frac{\gamma_{\text{cav}}}{\gamma_{\text{fs}}} &= 1 + \eta_{\text{QY}} DW F_{\text{P}}, & \text{(linear regime)} \label{eq:gamma-ratio},
\end{align}
with $\eta_{\text{QY}}=\gamma^\text{R}/\gamma_{\text{fs}}$.
Note that $\phi_{fs}$ is the total photon flux, including the phonon sidebands.

The emitter-cavity spectral matching is optimized using a longitudinal sweep of the cavity length (Fig.~\ref{fig:PL}~(b)). The resonance appears as a bright spot at the crossing of the cavity dispersion and the emitter emission energy. Setting the cavity length at resonance, we record the emission spectrum shown in Fig.~\ref{fig:PL}~(c). Note that the excitation power may be slightly modified  in the cavity due to interferences arising from the finite reflectivity of the mirrors at 1064 nm, which prevents direct comparisons of the count rates in the linear regime. However, measurements at saturation are immune to this effect. 

\begin{figure}[ht]
\includegraphics[width=8.6cm]{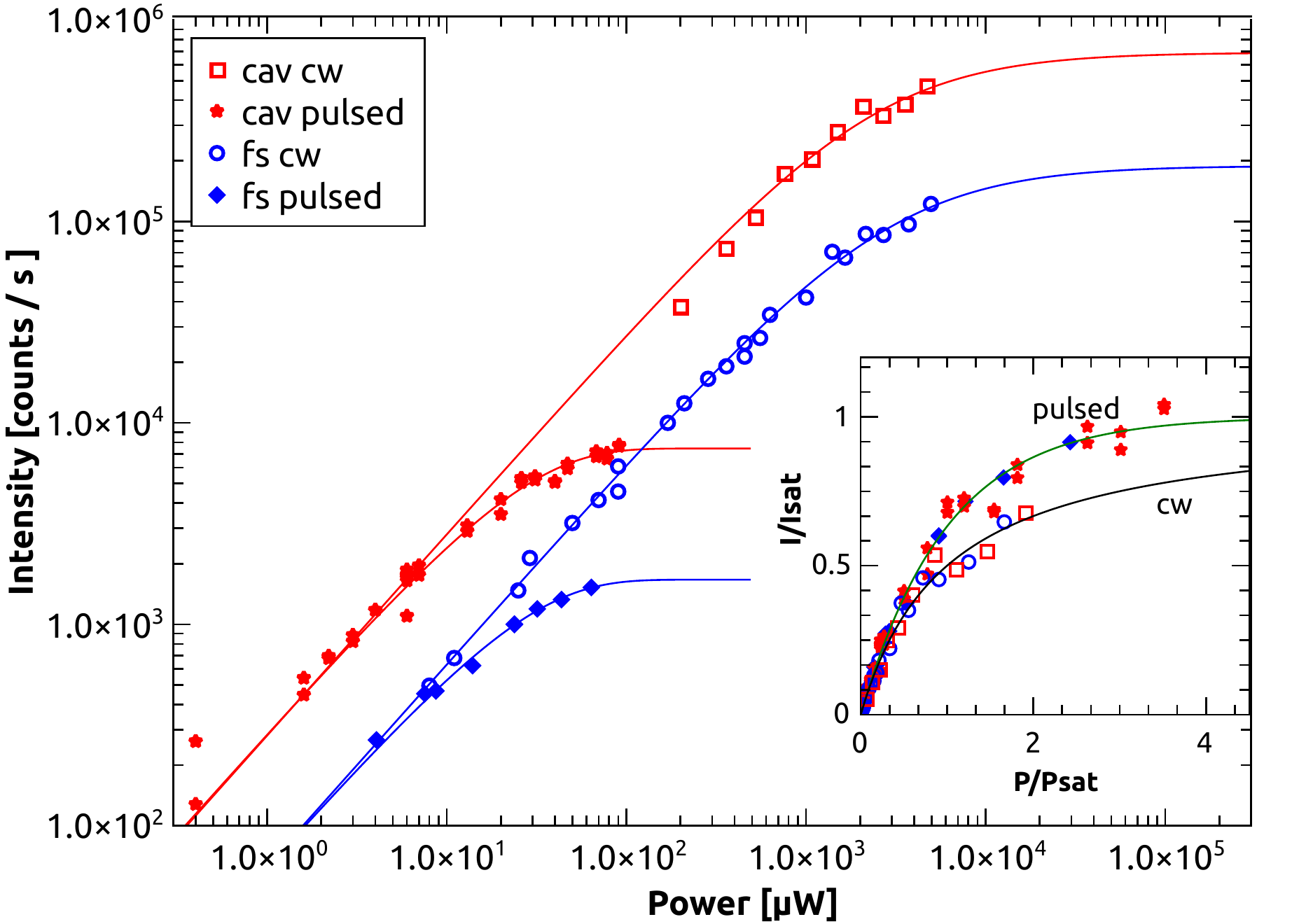} \caption{Spectrally integrated emission count rates as a function of the excitation power (log-log scale) for the FS configuration (blue symbols) and cavity configuration (red symbols) for cw excitation (open symbols) and pulsed excitation (full symbols). 
(The excitation power for the FS pulsed line has been rescaled by a factor 0.59 so that the low linear part of the curve matches the cw one (excitation misalignment).  Inset : same data in linear scale. \label{fig:satur}}.
\end{figure}

We first performed saturation measurements on the same nano-emitter in free space and in cavity at the lowest reachable mode volume (Fabry-Pérot mode $p=6$, $p$ being the longitudinal order, \textit{i.e.} the cavity length in units of $\lambda/2$ which can be extracted from the free spectral range), under pulsed or cw excitation at an excitation wavelength of 1064 nm. Fig.~\ref{fig:satur} shows the raw count rates obtained after integration over the whole spectrum as a function of the excitation power in FS and in cavity (flat mirror port) configurations. The saturation plateau is higher in cavity than in free space, which is a direct signature of the brightness enhancement. 

After careful calibration of all optical parts of the setup (see SI), we show that, at resonance, one count on the CCD corresponds to $41$ photons exiting the cavity from the flat mirror port and simultaneously $44$ photons emitted in the fiber port. At saturation, the detector count rate ($4.7\times 10^5 $ counts/s) thus corresponds to a single mode fiber coupled photon flux of $2.1\times10^7$ photons/s. Compared to the FS configuration, this flux corresponds to a brightening of the source "in the first lens" of about 16.

To quantify the genuine emission enhancement (that is, at the level of the emitter) and hence the Purcell factor, we need to correct the previous count rates from the extraction efficiencies for both cavity outputs and for the FS configuration. To this aim, we used numerical simulations (see SI). 
With very similar dielectric coatings on the fiber and on the flat mirror, the exit probabilities are expected to be rather balanced. However, due to the high quality factor reached in this cavity, the absolute exit probabilities are extremely sensitive to any small internal losses (scattering in the dielectric stack due to residual roughness, in the polystyrene layer...), which constitute pure losses (in contrast to the transmission losses which are "useful losses"). In order to assess these pure losses, we compared the simulated and measured Q factors of the cavity as a function of the cavity length within the cavity stability range (cavity modes from $p=6$ to $p=9$).
 In particular, due to the pencil-shape of the fiber apex, additional spill-out losses are observed in the simulations that yield an almost constant quality factor for all the modes $Q_{\text{cav}}\simeq5.5 \pm 0.5\times10^4$ (see SI). In addition, the numerical simulations showed that the photons exiting through the optical fiber do not perfectly couple to the fiber core but can be lost in the cladding due to the important  curvature of the mirror. Experimentally, we measured the cavity quality factor at the emitter wavelength by ring-down measurements and obtained a constant value $Q_{\text{cav}}\simeq 1 \pm 0.1 \times 10^4 $ for all the longitudinal modes from $p=6$ to $p=9$. The significant mismatch between the simulated and experimental values is assigned to the internal losses, which amount to 1300 ppm. Then, we deduce effective exit probabilities of 7.8\% through the flat mirror and 6.0\% through the fiber mirror for the cavity mode $p=6$. \footnote{If the origin of these internal losses could be identified and suppressed, we could expect a coupling as high as 31\% into the single mode fiber and 40\% through the flat mirror, assuming that the losses towards the cladding are unchanged.} 
We deduce the ratio of the collection efficiencies in FS and in cavity configurations (flat mirror output) to be $\eta_{\text{fs}}/\eta_{\text{cav}}=4.9 \pm 0.2$. We thus obtain $\phi_{\text{cav}}/\phi_{\text{fs}} = 19 \pm 2$ at saturation, which yields from eq.~\ref{eq:sat-rates} and $DW=0.65\pm0.05$ a first estimate of the Purcell factor $F_\text{P}=29\pm4$. Note that the count rate ratio is the same (within the error bars) in the linear regime, which is consistent with $\eta_{\text{QY}} DW F_\text{P} \ll1$, as will be confirmed using lifetime measurements.




\section{Lifetime measurements} 

\begin{figure}[ht]
\includegraphics[width=8.6cm]{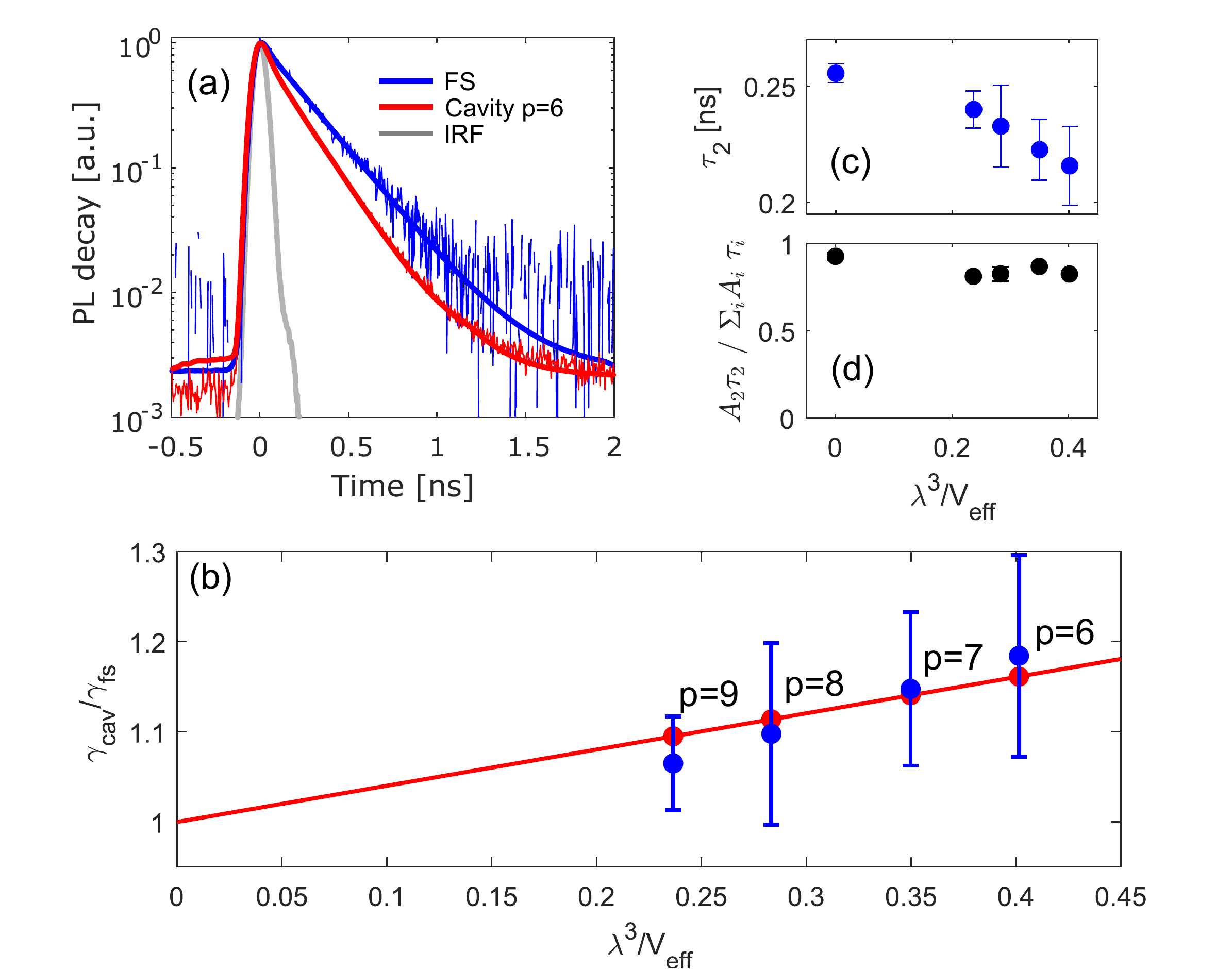} \caption{(a) Free space and cavity (p=6, $\lambda^3/V_{\text{eff}}=0.4$) lifetime comparison and their respective biexponential fits. The IRF is represented in grey line. (b) Experimental cavity decay rate acceleration factor (blue open circle) as a function of $\lambda^3/V_{\text{eff}}$ and the fit according to equations \ref{eq:purcell_theo} and \ref{eq:gamma-ratio}. The error bars are calculated from repeated lifetime measurements. (c) Lifetimes as a function of the inverse of the cavity mode volume. (d) Relative contribution of the long time component extracted from the biexponential fits over the signal integral $A_2 \tau_2/(A_1 \tau_1 + A_2 \tau_2)$. \label{fig:Life}}
\end{figure}

Comparison of the lifetimes in FS and in cavity configurations brings a direct insight into the Purcell acceleration of the radiative decay.  
Time-resolved measurements were performed using superconducting photodetectors (SSPD) coupled to a PicoHarp 400 aquisition card. Spectral selection was ensured by  dichroic bandpass filters.
The PL transient in FS shows a biexponential behavior with a short time $\tau_1=23 \pm 5$~ps and a long time $\tau_2=256\pm 4$~ps (Fig.~\ref{fig:Life}~(a)). 
At the smallest reachable cavity mode ($p=6$), a biexponential fitting of the transient yields $\tau_{2}^{\text{cav}}=216 \pm 15$~ps and $\tau_{1}^{\text{cav}}=28 \pm 2$~ps. The relative weight of the long component ($A_2\tau_2/(\sum A_i\tau_i)$) being larger than 80\%, we define the lifetime ratio as  $\tau_{2}^{\text{fs}}/\tau_{2}^{\text{cav}} = 1.19 \pm 0.09$. We repeated such lifetime measurements for different mode volumes. The data is shown in Fig.~\ref{fig:Life}~(c-d) as a function of $\lambda^3/V_{\text{eff}}$.  For each mode this mode volume was computed by finite element simulation (see SI). The emitter lifetime decreases as the inverse of the cavity mode volume in agreement with theory (eq. \ref{eq:purcell_theo} and \ref{eq:gamma-ratio}), as shown in Fig.~\ref{fig:Life}~(b).  

Using the count rates and recombination rates ratios $\phi_{\text{cav}}/\phi_{\text{fs}}$ and $\gamma_{\text{cav}}/\gamma_{\text{fs}}$ we are able to extract both the Purcell factor $F_\text{P}=29\pm4$ (in line with the estimate based on brightening of the previous section) and the emitter quantum yield $\eta_{\text{QY}}=1\pm0.4\ \%$. This value is confirmed by a direct experimental estimate. In fact, in pulsed FS saturation measurements, the saturation intensity reads $I_{\text{sat}}=\eta_{\text{coll}} \eta_{\text{QY}}  f_{rep}$, where $\eta_{\text{coll}}$ is the collection efficiency of the optical setup in free space, $f_{\text{rep}}$ is the pulse laser repetition rate. Hence, we deduce $\eta_{\text{QY,sat}}=0.7\pm0.2\ \%$.
This radiative quantum yield  is relatively low compared with value reported at room temperature \cite{Piao2013}, but similar to the values reported at low temperature for intrinsic (non functionalized) nanotubes. This effect, which has to be confirmed on a larger data set, could be related to a dark state trapping the population at cryogenic temperatures.  
\newline

\section{Phonons assisted CQED effects}


Interestingly, the FS spectrum $S_{\text{NT}}(\omega)$ of the emitter deviates significantly from a Lorentzian with side-bands that result from acoustic phonon assisted emission processes \cite{Vialla2014a}. When the cavity is detuned from the ZPL, the output signal remains sizeable over a significant spectral range, which results from the coupling of the emitter to the cavity through phonon assisted processes.  
A factor of merit of the tuning capability of our system is the spectral profile of the brightness  $\beta(\omega_{\text{cav}})$,  which describes the emission probability in the cavity mode for each single excitation of the emitter as a function of the cavity resonance frequency $\omega_{\text{cav}}$. 

The expression of $\beta(\omega_{\text{cav}})$ has been derived in \cite{chassagneux2018} as a function of the spectral and dynamical properties of the emitter in free space and reads :
\begin{equation}
\beta(\omega_{\text{cav}}) = \frac{g^2\Tilde{S}^{\text{emi}}(\omega_{\text{cav}}) /\gamma}{1+g^2 \tilde{S}^{\text{emi}}(\omega_{\text{cav}})/\gamma+g^2 \tilde{S}^{\text{abs}}(\omega_{\text{cav}})/\kappa},
\label{eq:beta}
\end{equation} 
where $\tilde{S}^{emi}(\omega_{\text{cav}})=(S_{\text{NT}}*\mathcal{L})(\omega_{\text{cav}})$ is the FS emission spectrum convoluted to a Lorentzian profile $\mathcal{L}(\omega)=[(\pi\kappa/2)(1+(2\omega/\kappa)^2)]^{-1} $.  It represents the overlap of the cavity mode with the FS emission spectrum (cf SI). The relationship with the usual Purcell picture is that $\frac{\gamma_{\text{cav}}}{\gamma_{\text{fs}}}=1+g^2 \tilde{S}_{\text{max}}/\gamma_{\text{fs}}$, which connects $\tilde{S}_{\text{max}}$ to the Debye-Waller factor $DW$ factor through $\tilde{S}_{\text{max}} = 4 DW /(\gamma^*+\kappa)$.

Note that in our system $\kappa \simeq 30 \gamma$ so that the re-absorption contribution $g^2 \tilde{S}^{\text{abs}}/\kappa$ can be neglected. The dynamical parameters $\gamma$ and $\kappa$ are deduced from time-resolved measurements. Hence, we can in principle deduce the value of the Rabi coupling $g$ from measurements of the brightness spectral profile.

\begin{figure}[ht]
\includegraphics[width=8.6cm]{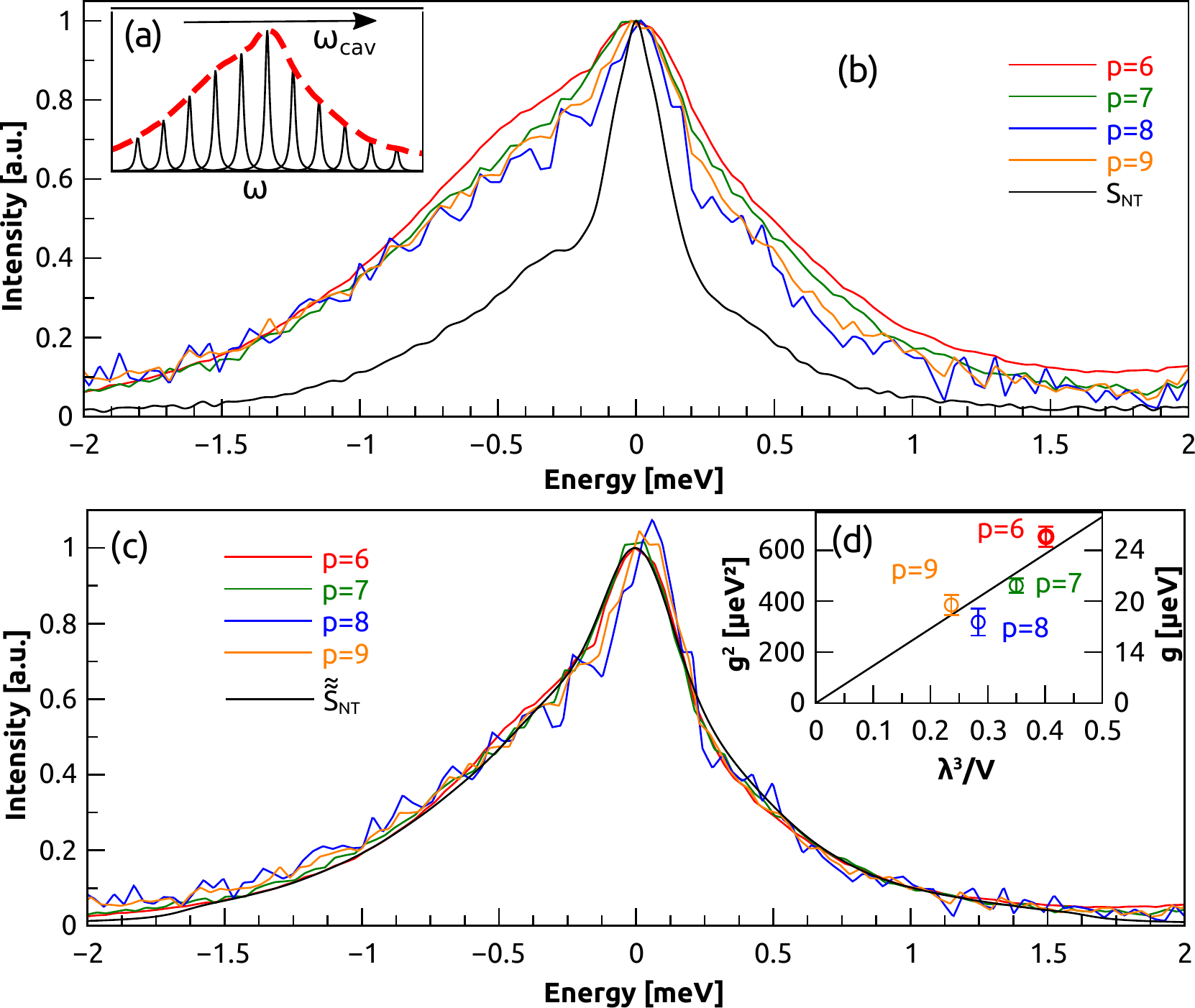} \caption{(a) Principle of the measurement of the brightness spectral profile: the envelope of the output spectrum is measured as a function of the cavity energy $\omega_{\text{cav}}$, either point by point for each cavity detuning (black plain lines) or dynamically by modulating the cavity frequency resonance $\omega_{cav}$. The latter method is used in the following. The normalized envelope $\text{E}_{\text{mod}}$ is depicted in red; it is closely related to $\beta$. (b) Experimental normalized envelopes $\text{E}_{\text{mod}}$ for several longitudinal cavity modes and compared with the free space spectrum  $S_{\text{NT}}$ (black). (c) Spectral profiles obtained after numerical inversion of $\text{E}_{\text{mod}}$ to retrieve $\stackrel{\approx}{S}_{\text{NT}}$ with $g^2/\gamma$ as free parameter (cf SI). (d) Values of $g^2$ (resp. $g$, right nonlinear scale) deduced from the previous analysis as a function of $\lambda^3/V_{eff}$ together with a linear fit (grey)  \label{fig:Pmod1}}
\end{figure}

To finely measure $\beta(\omega_{\text{cav}})$ despite mechanical vibrations and spectral diffusion, 
we chose to modulate the cavity detuning (at a frequency of 65 Hz) over a large spectral range in order to obtain directly the envelope of the output spectrum $\text{E}_{\text{mod}}(\omega)$ as shown in Fig.~\ref{fig:Pmod1}~(a) and (b).  The coupling to the cavity yields a drastic enhancement of the relative weight of the phonon sidebands compared to the ZPL, enlarging considerably the working spectral range of the single-photon source compared to its free space bandwidth.  This envelope profile is closely related to the brightness spectral profile and in the limit of an incoherent pumping of the cavity by the nanotube, the relation reads (cf SI and \cite{Roy-Choudhury2015}): ${\text{E}_{\text{mod}}(\omega)=c (\beta * \mathcal{L})(\omega) \simeq c\  a\stackrel{\approx}{S}_{\text{NT}}/(1+a\stackrel{\approx}{S}_{\text{NT}})}$, where $c$ is a constant, $a=g^2/\gamma$ and the double tilde notation refers to a double convolution with $\mathcal{L}$. 
Here, the cavity spectrum is five times narrower than that of the emitter, hence the convolution with $\mathcal{L}$ has a small effect but cannot be completely neglected.     

To rule out any spurious broadening effect, we took advantage of the fact that the expression of $\text{E}_{\text{mod}}$ can easily be inverted, so that the free-space emission spectrum can be recovered from the envelope modulation profile: ${\stackrel{\approx}{S}_{\text{NT}} \simeq \text{E}_{\text{mod}} / (a\  c- a\ \text{E}_{\text{mod}})  }$. 
Hence, we can perform this simple algebraic transformation for each longitudinal mode with $g$ as the only free parameter. The outcome of this transformation is displayed in Fig.~\ref{fig:Pmod1}~(c) where the transformed envelope modulation profile obtained for each longitudinal mode nicely falls back on the free-space spectrum convoluted by the cavity profile $\stackrel{\approx}{S}_{\text{NT}}$, proving the cQED origin of the apparent spectral broadening. The inset of Fig.~\ref{fig:Pmod1}~(c) shows the values of $g^2$ as a function of the inverse of the mode volume for each longitudinal mode, together with a linear fit. The values obtained for the smallest mode volumes reach up to $25\;\mu$eV. This linear dependence is in agreement with theory and simply reflects the increasing fluctuations of the vacuum electric field when reducing the mode volume and hence the increasing dipolar coupling. Compared to previous reports \cite{Jeantet2017a}, we obtain a significantly larger Rabi splitting (up to a factor of 4), which mostly reflects that the ratio $\lambda^3/V$ mechanically increases when shifting the working wavelength to the telecom bands while keeping a similar geometrical mode volume.

 In principle, the three cQED effects investigated in this work (brightening effect, lifetime reduction and spectral enlargement of the brightness pofile) reflect similarly the enhancement of light-matter coupling encoded in $g$. However, deriving a quantitative relationship between these quantities is not straightforward due to the interplay between spectral diffusion and pure dephasing in the linewidth of the emitter \cite{Jeantet2017a}. For instance, when neglecting spectral diffusion with respect to pure dephasing in a first approach, the $g$ values deduced from the lifetime measurements ($2g=\sqrt{\gamma^*\Delta \gamma/DW}$) are consistently smaller than that deduced from the spectral method (by a factor of $\sim 3$), showing that fast spectral diffusion remains a significant contribution to the linewidth even for the thinnest ZPL profiles ($\sim 200 \mu$eV). This is consistent with the strong bunching effect observed in the intensity correlation measurements (Fig.~\ref{fig:PL}~(e)) on a time-scale of a few nanoseconds \cite{Walden-Newman2012}. 
Hence, we emphasize that the three approaches presented here are complementary and yield accurate estimate of their respective cQED figures of merit ($\Phi/\Phi_0$, $F_\text{P}$ or $g$) which are connected to each other in a non trivial way in complex condensed-matter emitters combining non-radiative recombination, pure dephasing and spectral diffusion.

\section{Conclusion}

In conclusion, we have coupled a single organic colored center grafted on a carbon nanotube to a tunable fiber microcavity, achieving a bright, telecom wavelength tunable single-photon source, with a fibered output count rate up to 20 MHz. Using the unique versatility of our setup we could explore independently different figures of merit of cQED effects. We demonstrated a brightening effect of about 20 in the first lens, even at low pumping rate, which reflects the strongly beneficial Purcell effect for dim emitters (low quantum yield). Time-resolved measurements have brought a direct insight into the acceleration of the lifetime of the emitter, yielding independent estimates of the Purcell factor ($F_\text{P} \simeq 30$ ) and the intrinsic quantum yield of the emitter at low temperature which remains surprisingly low for a colored center. Finally, using the original open geometry of the setup, we could tune the working wavelength over a broad spectral range and observe cQED effects on the phonon side-bands of the emitter. We observed a strong broadening and warping of the brightness spectral profile, which is a unique signature of cavity enhanced phonon assisted emission mechanisms. This approach yields an independent estimate of the Rabi splitting of the system  (up to $25\mu$eV). This strong enhancement of light-matter interaction  paves the way to reaching the strong coupling regime with a single emitter, a requirement for many advanced quantum technology applications.

\onecolumn
\section{Supplementary informations}

\subsection{Cavity characteristics}

\begin{figure}[h!]
    \centering
    \includegraphics{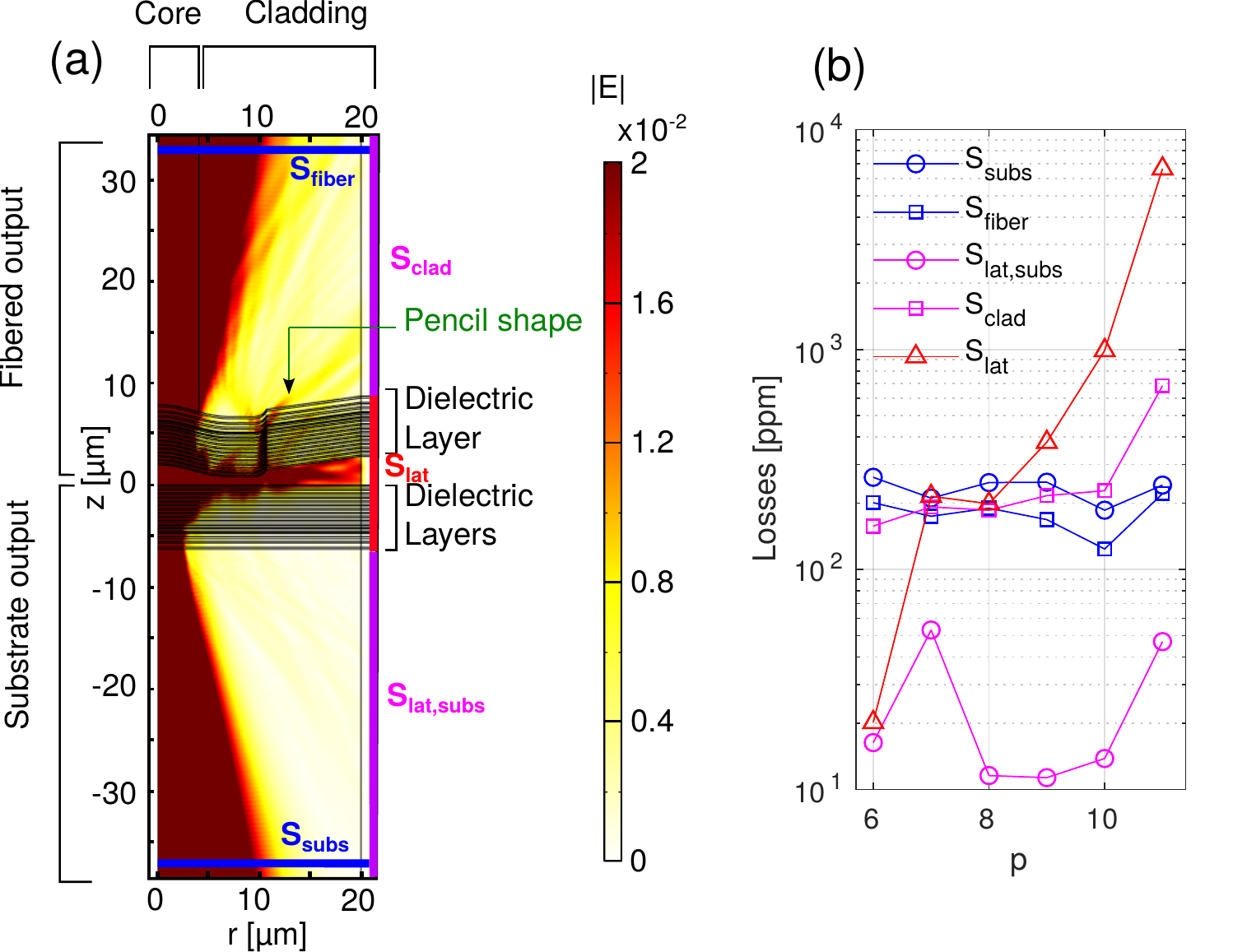}
    \caption{Finite element numerical simulation of the electric field, $|E|$ in the cavity, computed at a wavelength of 1.275~$\mu$m (a). The blue, red and magenta lines represent the surfaces through which the losses are computed and reported in (b) as a function of the longitudinal mode order "p".}
    \label{fig:Simul_losses}
\end{figure}

We measured the geometry of the concave micro-mirror with an AFM. We simulated the electric field of the cavity in cylindrical coordinates. In figure \ref{fig:Simul_losses}~(a), the dielectric interfaces are represented in black solid lines. The Bragg mirrors consist in alternating high and low refractive index layers resulting in a stop band in the range 1.25 µm to 1.55 µm. The top layer of the fiber is a high refractive index layer made of $TiO_2$, and the top layer of the planar mirror is the of low refractive index $SiO_2$. The polystyrene layer on top of the planar mirror is also considered.

The electric field magnitude reaches a maximum of 58 in arbitrary units. In order to visualize the losses channels of the system, we use a saturated $|E|$ colour scale (max at $10^{-2}$ arbitrary units). The losses are computed through the flux across the surfaces shown in blue, red and magenta in figure \ref{fig:Simul_losses}.(a). The losses were computed for several Fabry-Pérot longitudinal modes ranging from 6 to 11 at a wavelength of 1.275 µm, as shown in figure \ref{fig:Simul_losses}~(b).

We observe that spill-out losses are no more negligible as soon as \(p>6\), which may be due to a tip effect of our pencil shape geometry. They even overcome the other losses by an order of magnitude when \(p>9 \)  leading to a restricted cavity stability length of approximately 6 µm, significantly lower than the theoretical value given by the radius of curvature ($\mathcal{R}=10$ µm) of the concave mirror. In addition we observe losses occurring through the cladding of the fiber, which are comparable in magnitude to the collection channels $S_{subs}$ and $S_{fiber}$, meaning a substantial share of photons are lost through the cladding. Using these simulations, we computed the effective mode volume and the quality factor. These computed values $Q_{cav}$ are reported in Table \ref{tab:Comparaison_Q} and compared to the experimental ones obtained by ring-down measurements.

The difference between the computed and the measured quality factors is attributed to residual scattering/absorption losses arising from the PS layer (eg : due to the surface roughness, impurities...). We assume these losses are independent of the mode order. Finally, taking into account all the losses channels we computed the exit probabilities for the photons emitted by the nanotube either through the fiber or through the back mirror as a function of the cavity mode order as shown in Table \ref{tab:Comparaison_Q}. 


        

\begin{table}[h!]
    \centering
    \begin{tabular}{|c|c|c|c|c|c|}
    \hline
        p & $V_{eff}$ ($\lambda^3$) & $Q_{cav,th} (\times10^4)$ & $Q_{cav,exp} (\times10^4)$ & $P_{cav\rightarrow subs,exp} (\%)$ & $P_{cav\rightarrow fiber,exp} (\%)$    \\
        \hline
        6 & 2.49 & 5.69 & 1.12 & 7.85 & 6.03  \\
        7 & 2.86 & 4.92 & 1.05 & 5.32 & 4.40 \\
        8 & 3.53 & 6.02 & 0.99 & 4.89 & 3.74 \\
        9 & 4.23 & 5.45 & 1.04 & 4.64 & 3.14\\
        \hline
        
    \end{tabular}
    \caption{Mode volume,$V_{eff}$, computed cavity quality factor $Q_{cav,th}$ and measured quality factor $Q_{cav,exp}$ as a function of the longitudinal mode order $p$. Exit probabilities for the  photons through the substrate ($P_{cav\rightarrow subs,exp}$) , or through the fiber $P_{cav \rightarrow fiber,exp}$. The total differs from 100\% due to losses. }
    \label{tab:Comparaison_Q}
\end{table}

\subsection{Assessment of optical losses in the optical paths}

The optical transmission coefficients of the optical parts or collection/detection efficiencies of the instruments were either measured directly using a 1.31~µm laser diode for the lenses and mirrors, or estimated numerically for the collection of the custom objective in free-space configuration and for the output probabilities of the cavity (see Table~\ref{tab:Comparaison_Q}). They are reported in Table \ref{tab:Optical efficicencies}. The extraction efficiencies take in account the numerical aperture of the objectives and the radiation diagram of the emitter both in cavity and free space. 

In cavity, the photon beam exiting through the planar mirror is collimated by a 0.16 numerical aperture aspheric lens with a 5~mm diameter. In addition, a plano-convex lens is installed onto the back of the 6~mm thick planar mirror to improve the collection by the aspheric lens. However, these optical parts being inside the cryostat, their position cannot be finely adjusted and the effective collection efficiency has to be calibrated empirically. 

    

\begin{table}[!ht]
    \centering
    \begin{tabular}{|c|c|c|c|}\hline
         & Free-space & Cavity : planar mirror port (p=6) & Cavity : fiber port (p=6)   \\
        \hline
        Extraction in the 1st lens& 19\% & 5.6\% & 6.0\% \\ \hline
        Beamsplitter & 61.2 \% & 98 \% & -- \\
        Optical parts in the cryostat & -- &34 \%& -- \\
        Other parts along the path & 58\% & 74\% & 19\% \\
    
        \hline
        Spectrometer + CCD & 9.8\% & 9.8\% & -- \\
        \hline
        SSPD & -- & -- & 78\% \\
        \hline
    \end{tabular}
    \caption{Extraction yield "in the first lens" (or in the collecting fiber) estimated from numerical simulations. Transmission coefficient of optical parts along the collection path for free-space and cavity configurations deduced from measurements at $\lambda=1310$ nm. Efficiency specifications for the spectrometer + CCD detector and for the SSPD detector.  }
    \label{tab:Optical efficicencies}
\end{table}

To this aim, we recorded simultaneously the flux of the emitter at 1.275 µm in the cavity mode (p=6) with the spectrometer for the planar mirror output and with the SSPD for the fiber output with an cw excitation power of 150 µW. The count rates are shown in Figure~\ref{fig:Comparison_flux}. Using Table \ref{tab:Optical efficicencies}, we can experimentally evaluate the ratio of photons exiting the cavity through the fiber over the one exiting the cavity through the planar mirror. The experimental ratio is 2.3, which means, using the values of Table~\ref{tab:Optical efficicencies}  that the collection efficiency of the optical parts in the cryostat is about 0.33. 
Finally, using this calibration we estimate the collection efficiency ratio for the free-space vs cavity mode through the planar mirror : {$\eta_{\text{fs}}/\eta_{\text{cav-plan}}=4.9$}. 

    

\begin{table}[h!]
    \centering
    \begin{tabular}{|c|c|c|c|c|}\hline
         & Free-space & Cavity (p=6) &  Cavity (p=6) &     \\
         &&planar mirror port & fiber port &\\
        \hline
        Extraction in the 1\textsuperscript{st} lens& 19 \% & 5.6 \% & 6.0 \% & simulations \\ \hline
        Transmission along the path &&&&\\ and detector efficiency & 3.5 \% & 2.4 \% & 15 \% & experimental\\
        \hline
        Overall efficiency & 0.66 \% & 0.135 \%& 0.90 \%&  \\    
        \hline
    
    \end{tabular}
    \caption{Summary of useful extraction and overall collection efficiencies in all the configurations of the experimental setup.}
    \label{tab:Optical efficicencies2}
\end{table}

\subsection{Single mode photon flux in the single mode fiber}
\begin{figure}[h]
    \centering
    \includegraphics{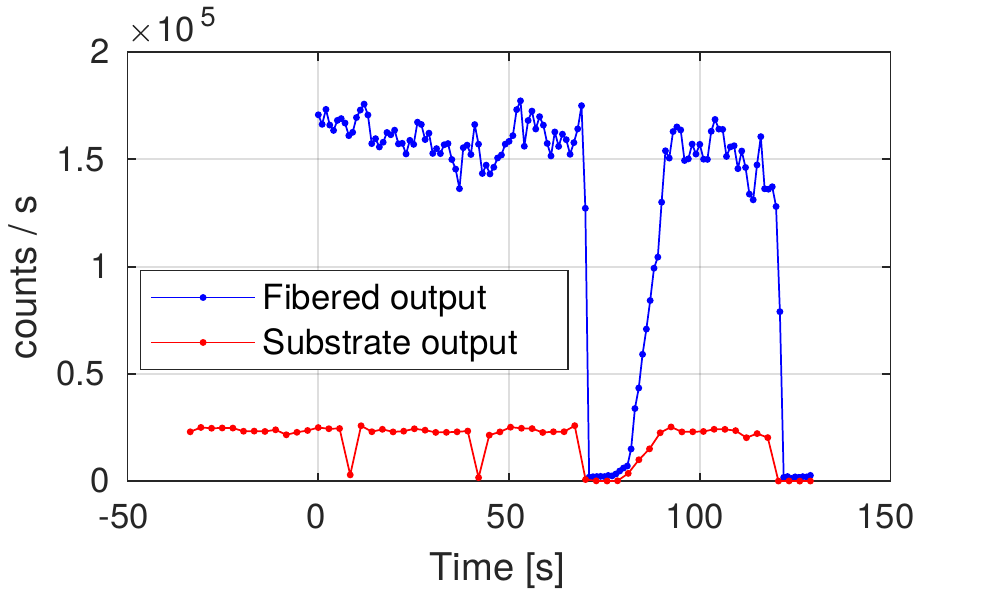}
    \caption{Cavity coupled photon flux recorded as a function of time either from the fiber output or from the planar mirror. The ratio of the fibered output to the planar output is $\simeq 6.7$} 
    \label{fig:Comparison_flux}
\end{figure}

To extract the single photon flux in the single mode fiber, we record at the same time the count rate of photons with the SSPD and the spectrum of the photons collected through the planar mirror with the spectrometer and CCD (integration time 1~s), with 100~µW cw excitation power. 
The count rates from each output port are shown in Figure~\ref{fig:Comparison_flux}. We observe a ratio of $6.7\pm0.3$ between the flux recorded on the SSPD and the one measured on the spectrometer (after spectral integration over the linewidth).

The flux measured on the spectrometer is typically $2.7 \pm0.1\times10^4$counts/s, for an excitation power of 100~µW. We deduce from CCD measurements that the fiber-coupled flux at the maximum excitation power of 4.7~mW is $3.1\pm0.3\times10^6$ counts/s. Correcting from the optical efficiencies along the optical path towards the CCD, we deduce a typical flux of $21\pm2\times10^6$ photons/s coupled into the single mode fiber. 

\subsection{Extraction efficiency}

\begin{figure}[h!]
    \centering    \includegraphics[width=14cm]{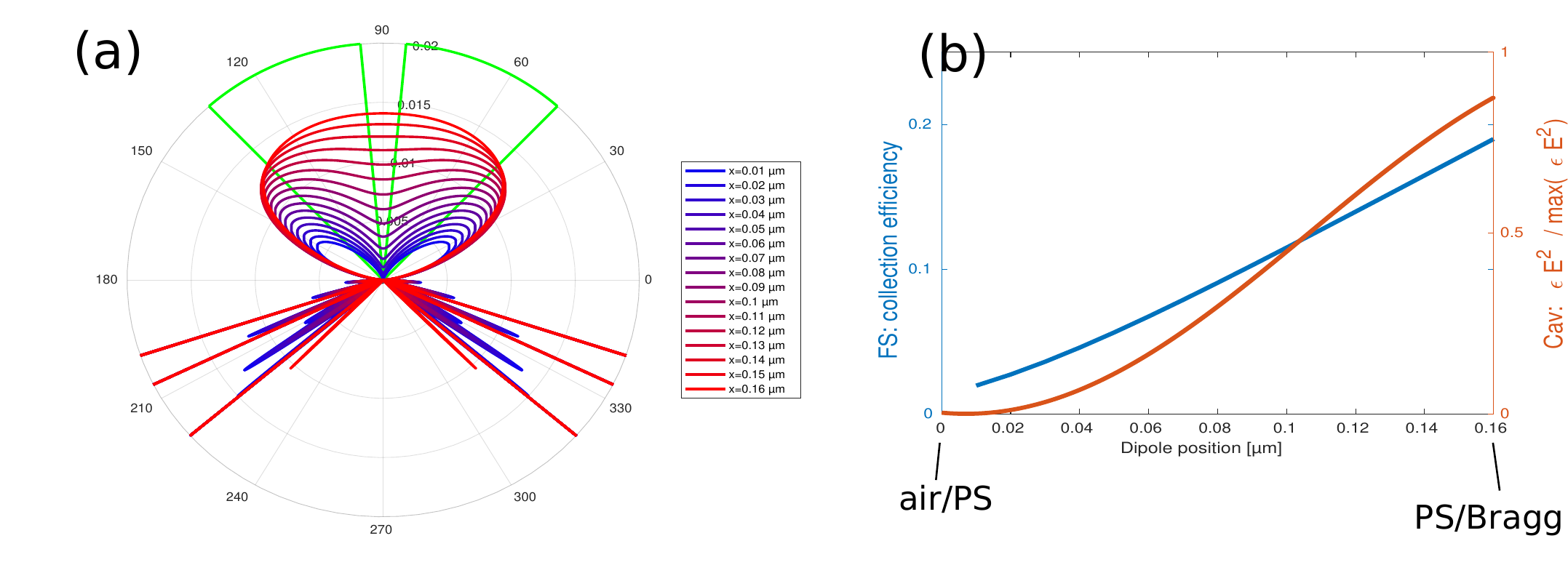}
    \caption{(a) Calculated FS emission diagram\cite{Reed1987}  for different position of a dipole in a 160nm thick polystyrene layer deposited on a Bragg mirror. The $x=0$ position corresponds to the air-PS interface and $x=0.16 \mu m$ to the PS-Bragg interface. The green pattern indicates the angle collected by the drilled aspheric lens. (b) Comparison of the FS collection efficiency and the ratio $\epsilon E^2 / \text{max}(\epsilon E^2)$ which corresponds to the variation of the emitter-cavity coupling $g^2$ as a function of the vertical position of the emitter in the PS layer.}
    \label{fig:emissiondiagram}
\end{figure}

In FS configuration, the emission diagram was computed numerically \cite{Reed1987} for several locations of the emitter in the polystyrene layer and the collection efficiency was computed using the numerical aperture of the aspheric lens (N.A.=0.7). For a polystyrene layer thickness of 160 nm, the collection efficiency varies from 19 \% if the emitter is close to the Bragg mirror down to 2 \% if the emitter is close to the polystyrene surface due to interferences with the PL reflected off the Bragg mirror. This effect yields a selection bias as we mainly  chose the brightest emitters. Fortunately, the interference pattern is the same in the cavity configuration, so that the brightest FS emitters are also those that are best spatially coupled to the cavity mode. Hence, as we selected only the brightest emitters, we assume that the the FS collection efficiency is 19 \%. In any case, any possible bias if the estimate of this collection efficiency would affect similarly the coupling to the cavity, resulting in a negligible modification in the estimate of the Purcell factor. 

\subsection{Coupling to phonons}
The coupling to phonons can be taken into account following our previous work \cite{chassagneux2018} and \cite{Roy-Choudhury2015}. In the polaron framework and in the interaction picture, by using a second order Born approximation, the master equation reads  \cite{chassagneux2018}:
\begin{equation}
\frac{\partial \rho_I}{\partial t} = -\frac{1}{ \hbar^2}\int_0^t du [V_I(t),[V_I(u),\rho_I(t)]],
\end{equation}
where, $V_I(t) = i\hbar g ( e^{-i\delta t} e^{i\Omega(t)}a^\dag \sigma^- - h.c.) $. We trace out the phonon bath degree of freedom, using $Tr_B(\rho_{th}e^{\mp i \Omega(t)}e^{\pm i \Omega(u)}) = K(t-u)$, where $K(t)$ is the phonon kernel. By letting $t$ go to $\infty$ in the upper limit of the integral, and by introducing a dephasing term, one finally gets:
\begin{equation}
 \frac{\partial \rho}{\partial t}  = \frac{1}{i\hbar} [H_{eff},\rho] + g^2 \tilde{S}^{\text{emi}}(\omega_{\text{cav}}) L[a^\dag\sigma^-]+ g^2 \tilde{S}^{\text{abs}}(\omega_{\text{cav}}) L[a\sigma^+],
\end{equation}
where $\tilde{S}^{\text{emi /abs}}$ are defined in \cite{chassagneux2018}, with the normalisation $\int d\omega\ \tilde{S}^{\text{emi /abs}}(\omega)=2\pi $. The Lindblad terms are given by: $L[C] = C\rho C^\dag- \frac{1}{2} (C^\dag C \rho +\rho C^\dag C)$.
The effective hamiltonian (in the polaron framework) is given by:
\begin{equation}
H_{eff} = \hbar (\omega_x+\Delta_x)  \sigma^+\sigma^- + \hbar (\omega_{c}+\Delta_{c})    a^\dag a
\end{equation}
where $\Delta_x$ and $\Delta_c$ are effective Lamb shift in the presence of phonons. 
The effective Hamiltonian is much simpler than the one in \cite{Roy-Choudhury2015}  since in a 1D phonon bath, the effective coupling $g' = g \langle B\rangle$ goes to zero.

Within this approximation the hamiltonian evolution of the exciton and the photon are independent and are coupled only through incoherent population transfer via the Lindblad terms.

\subsubsection{Emission spectrum}
The emission spectrum of the coupled nanotube-cavity system is given by :

\begin{equation}
S^{\text{NT in cav}} = \int_0^\infty dt \int_0^\infty d\tau e^{-i\omega \tau} \langle a^\dag(t+\tau) a(t) \rangle +cc   
\end{equation}
The expectation value of the two-time operator can be obtained  using the quantum regression theorem yielding :
\begin{equation}
\langle a^\dag(t+\tau) a (\tau) \rangle = e^{i (\omega_c+\Delta_c) \tau} e^{- \frac{\kappa+g^2 \tilde{S}^{\text{emi}}+g^2 \tilde{S}^{\text{abs}} }{2} \tau} \langle a^\dag(t) a(t) \rangle
\end{equation}

By neglecting the Lamb shift $\Delta_c$ and the additional cavity broadening  $\kappa+g^2 \tilde{S}^{\text{emi}}+g^2 \tilde{S}^{\text{abs}} \approx \kappa$ and under a weak continuous excitation of the emitter, the spectrum reads:
\begin{equation}
S^{\text{NT in cav}}  \propto \frac{1}{1+\left(\frac{\omega-\omega_{cav}}{\kappa/2}\right)^2} \langle a^\dag a \rangle_{ss}.
\end{equation}
The steady state $\langle\rangle_{ss}$ population is obtained by solving (in the steady state regime , with a weak incoherent pump $p$ ):
\begin{eqnarray}
\frac{d \langle \sigma^+ \sigma^- \rangle}{dt}  = 0 &=&  -(\gamma+g^2 \tilde{S}^{\text{emi}} )\langle \sigma^+ \sigma^- \rangle +  g^2\tilde{S}^{\text{abs}}\langle a^\dag a\rangle   + p , \\
\frac{d \langle a^\dag a\rangle}{dt}  = 0 &=&  -(\gamma+g^2 \tilde{S}^{\text{abs}} )\langle a^\dag a\rangle +  g^2\tilde{S}^{\text{emi}}\langle \sigma^+ \sigma^- \rangle    .
\end{eqnarray}
This leads to
\begin{equation}
\langle a^\dag a\rangle_{ss} = \frac{p}{k} \beta(\omega_{\text{cav}}),
\end{equation}
where \cite{chassagneux2018}
\begin{equation}
\beta(\omega_{\text{cav}}) = \frac{g^2\Tilde{S}^{\text{emi}}(\omega_{\text{cav}}) /\gamma}{1+g^2 \tilde{S}^{\text{emi}}(\omega_{\text{cav}})/\gamma+g^2 \tilde{S}^{\text{abs}}(\omega_{\text{cav}})/\kappa}.
\end{equation} 
Finally the emission spectrum is given by :
\begin{equation}
S^{\text{NT in cav}}(\omega,\omega_{\text{cav}})  \propto  \frac{2}{\pi \kappa}\frac{1}{1+\left(\frac{\omega-\omega_{cav}}{\kappa/2}\right)^2} \beta(\omega_{\text{cav}}).
\end{equation}
The equation can be interpreted in the following way : the Lorentzian factor, which is typically much narrower than the second factor, corresponds to the spectral filtering by the cavity. It yields an output spectral profile which is merely the Lorentzian transmission spectrum of the cavity. The second factor $\beta(\omega_{\text{cav}})$ corresponds to the overall amplitude of this Lorentzian signal for a given detuning $\delta$ between the cavity and the ZPL of the emitter.
By integrating the output spectrum over $\omega$ (which corresponds experimentally to measuring the output flux for a given detuning using a photo-diode), we recover $\beta(\omega_{\text{cav}})$. 

Reciprocally, if we modulate the detuning and measure the output spectrum over an integration time much larger than the modulation period, we obtain :
\begin{equation}
E_{\text{mod}}(\omega) = \int d\omega_{\text{cav}}  S^{\text{NT in cav}}(\omega,\omega_{\text{cav}}) = (\mathcal{L}*\beta)(\omega)   ,
\end{equation}
where $\mathcal{L}$ is a normalized Lorentzian profile $\mathcal{L}(\omega) =\frac{2}{\pi \kappa}\frac{1}{1+\left(\frac{\omega}{\kappa/2}\right)^2} $.

For the sake of simplicity, we verify numerically that :
\begin{equation}
(\mathcal{L}*\beta) \approx \frac{g^2\stackrel{\approx}{S} _{\text{emi}} /\gamma}{1+g^2 \stackrel{\approx}{S}_{\text{emi}}/\gamma+g^2 \stackrel{\approx}{S}_{\text{abs}}/\kappa}.
\label{approxdoubletilde}
\end{equation}
where the double tilde notation stands for a double convolution with $\mathcal{L}$. By using the coupling $g$ deduced from the following paragraph, we check numerically that the approximation (\ref{approxdoubletilde}) does not change the profile (the standard deviation of the left hand side minus the right hand side of (\ref{approxdoubletilde}) normalised to 1, is less than $5\cdot 10^{-5}$), but it boils down to a small vertical stretching of the profile by a factor $\approx 1.05$.

\subsubsection{Extracting the coupling $g$ from $\text{E}_{mod}$}

Naturally, the experimental modulation profile is known up to a global factor, hence we can only compare normalized profiles. Following this previous section, and by neglecting the re-absorption processes in (\ref{approxdoubletilde}), the normalized modulation profile reads :
\begin{equation}
    \text{E}_{\text{mod}}^\text{norm}(\omega)=\frac{1+a\stackrel{\approx}{S}_{max}}{a\stackrel{\approx}{S}_{max}} \frac{a\stackrel{\approx}{S}(\omega)}{1+a \stackrel{\approx}{S}(\omega)},
    \label{eq:norm_pofile}
\end{equation} 
where $\stackrel{\approx}{S}$ stands for $\stackrel{\approx}{S}_{emi}$, $\stackrel{\approx}{S}_{max}$ is the maximum value of $\stackrel{\approx}{S}$ and $a=g^2/\gamma$ is the parameter that will be extracted by fitting the normalized experimental profile to (\ref{eq:norm_pofile}).

\vspace{4cm}

\begin{acknowledgement}
The authors thanks X. He for sample preparation. The authors thank C. Diederichs, E. Baudin, G. Hetet, Z. Said for fruitful discussion as well as T. P\"opplau for fiber micro-machining. We also thank A. Leclercq and P. Morfin for help in designing and fabricating mechanical components. 
\end{acknowledgement}



\providecommand{\latin}[1]{#1}
\makeatletter
\providecommand{\doi}
  {\begingroup\let\do\@makeother\dospecials
  \catcode`\{=1 \catcode`\}=2 \doi@aux}
\providecommand{\doi@aux}[1]{\endgroup\texttt{#1}}
\makeatother
\providecommand*\mcitethebibliography{\thebibliography}
\csname @ifundefined\endcsname{endmcitethebibliography}
  {\let\endmcitethebibliography\endthebibliography}{}

\end{document}